\documentclass[12pt,a4paper]{article}

\usepackage{amsthm,amsmath,natbib,amssymb,amsfonts,bm}
\usepackage{graphicx}
\usepackage{placeins}

\usepackage[font=scriptsize]{caption}

\usepackage{subcaption}

\def\hang{\hangindent\parindent}
\def\rf{\par\noindent\hang}

\newtheorem{theorem}{Theorem}

\newtheorem*{theorem*}{Theorem}

\theoremstyle{definition}

\theoremstyle{remark}

\usepackage{moresize}

\def\hang{\hangindent\parindent}
\def\rf{\par\noindent\hang}

\newcommand{\REML}{\textsl{\ssmall REML}}
\newcommand{\ML}{\textsl{\ssmall ML}}

\begin{document}
	
	\baselineskip=20pt

\begin{center}
	{\bf \Large The effect of a Durbin-Watson pretest 
		on confidence intervals in regression}
\end{center}

%\bigskip

\bigskip

\begin{center}
	{\bf \large Paul Kabaila$^\textbf{*}$, Samer Alhelli, Davide Farchione and Nathan Bragg}
\end{center}

\medskip

\begin{center}
	%{\large
	{\sl Department of Mathematics and Statistics, La Trobe University, Australia}
	%}
\end{center}

	\bigskip

	\begin{center}
		\textbf{Abstract}
	\end{center}

  %\begin{abstract}
  	% No more than 150 words
  \noindent Consider a linear regression  model and suppose that our aim is to find a confidence interval for a specified linear combination of the regression parameters. In practice, it is common to perform a Durbin-Watson pretest of the null hypothesis of zero first-order autocorrelation of the random errors against the alternative hypothesis of positive first-order autocorrelation. If this null hypothesis is accepted then the confidence interval centred on the Ordinary Least Squares estimator is used; otherwise the confidence interval centred on the Feasible Generalized Least Squares estimator is used. We provide new tools for the computation, for any given design matrix and parameter of interest, of graphs of the coverage probability functions of the confidence interval resulting from this two-stage procedure and the confidence interval that is always centred on the Feasible Generalized Least Squares estimator. These graphs are used to choose the better confidence interval, prior to any examination of the observed response vector.
  	
  \vspace{0.5cm}
  	
  %\end{abstract}


  \vspace{1.5cm}

 \vspace{3cm}
 
 \noindent $^\textbf{*}$ Corresponding author. \textsl{E-mail address:} P.Kabaila@latrobe.edu.au

\newpage

  \section{Introduction}

Consider a linear regression model where the parameter of interest $\theta$ is a specified linear combination of the regression parameters.
Suppose that our aim is to find a confidence interval for $\theta$.
% that has minimum coverage probability $1-\alpha$. 
Commonly in econometrics, 
for example when the responses are measured over time,
the random errors in the regression model may be autocorrelated. In the absence of autocorrelation the usual confidence interval, centred on the Ordinary Least Squares (OLS) estimator and based on the assumption of independent errors, should be used. We call this the OLS confidence interval.
Of course, in the presence of autocorrelation, this 
interval is no longer valid. In this case, it is common to estimate the first order autocorrelation $\psi$, assuming that the random errors are a first order autoregressive process, and then to substitute this estimate into the expression for the confidence interval found using generalized least squares. We call this a feasible generalized least squares (FGLS) confidence interval.

The fact that the OLS confidence interval is preferable to the FGLS confidence 
interval when $\psi = 0$, has led to the proposal of the following two-stage procedure. We carry out a Durbin-Watson, or similar, pretest of the null hypothesis that $\psi =0$ against the alternative hypothesis that $\psi > 0$. 
%The Durbin-Watson test is implemented in a wide variety of software including 
% SAS, R, NCSS and  Excel.
If this null hypothesis is accepted then we use the OLS confidence interval; otherwise we use an FGLS confidence interval. We call this the two-stage 
confidence interval. This confidence interval
has been proposed by Wooldridge (2016, p.381), Kennedy (2008, p.119),
Anselin (2006, pp 931-2), Verbeek (2004, p.101), Berthouex and Brown (2002, pp 368-9), Salvatore and Reagle (2002, p.208), Giles and Giles (1993),  Pokorny (1987, pp.202-7), Folmer (1988), Griffiths and Beesley (1984), Katz (1982, pp.122-5) and Karmel and Polasek (1977, p.355).

A problem with the two-stage procedure is that the pretest may 
incorrectly accept or reject the null hypothesis, leading to a degradation in the coverage performance of the two-stage confidence interval. 
An alternative to the two-stage confidence interval is to always use a FGLS confidence interval. There are good reasons for constructing the FGLS confidence interval using the restricted maximum likelihood estimator (REML) of $\psi$, see  Cheang and Reinsel (2000). We will therefore construct the FGLS confidence interval using the REML estimator of $\psi$.

Our aim is to compare the coverage probabilities of the two-stage and FGLS confidence intervals. It would be nice if one could make some general statement,
such as ``the FGLS confidence interval always has better coverage properties 
than the two-stage confidence interval''. Our finding, however, is that this comparison depends crucially on the design matrix for the linear regression under consideration. We have therefore chosen to provide the tools for the comparison 
of the coverage probabilities for any given design matrix and parameter of interest. This comparison must,
of course, be carried out prior to any examination of the 
observed response vector.
In summary, the decision as to whether one uses the two-stage or FGLS confidence interval is made on a case-by-case basis, depending on the design matrix 
and parameter of interest at hand.

For simplicity, we have assumed that the random errors are an AR(1) process. As noted in the Remarks section, our methodology is easily extended to the case that the random errors are an ARMA$(\ell_1, \ell_2)$ process for any given $\ell_1$
and $\ell_2$ ($\ell_1 + \ell_2 >0$). In this section we also point out that our methodology can be easily extended to the case that the Durbin-Watson pretest is replaced by the so-called ``t-statistic''. Finally, we make some remarks about taking account of 
possible misspecification of the model for the random errors and 
an alternative framework for the construction of confidence intervals
for $\theta$ 
%and the analysis of their performance 
that 
accounts explicitly for this misspecification. This alternative framework includes the heteroskedasticity and autocorrelation consistent (HAC) estimator of the correct covariance matrix of the OLS estimator.

The main result of Section 2 is that the coverage probability of the 
FGLS confidence interval does not depend on either the regression parameters
or the variance of the random error. 
Consequently, for given design matrix, parameter of interest and nominal coverage, the coverage 
probability of the FGLS confidence interval is a function of $\psi$. 
The main result of Section 3 is that
the coverage probability of the two-stage confidence interval does not depend on either the regression parameters
or the variance of the random error. In fact, for given design matrix, parameter of interest, nominal  coverage and 
level of the Durbin-Watson pretest, the coverage probability 
of the two-stage confidence interval is a function of $\psi$.
This makes it easy to compare the coverage probabilities of the two-stage and
FGLS intervals for any given design matrix, parameter of interest, nominal coverage and level of the Durbin-Watson pretest. We estimate these coverage
probabilities using the variance reduction methods described in Section 4 and the Supporting Information, so that this comparison
can be quickly carried out.

\begin{figure}[]
	\begin{subfigure}{0.5\textheight}
%\begin{subfigure}
		%	\begin{subfigure}{1\textwidth}
		\centering
				\includegraphics[width=1.2\linewidth]{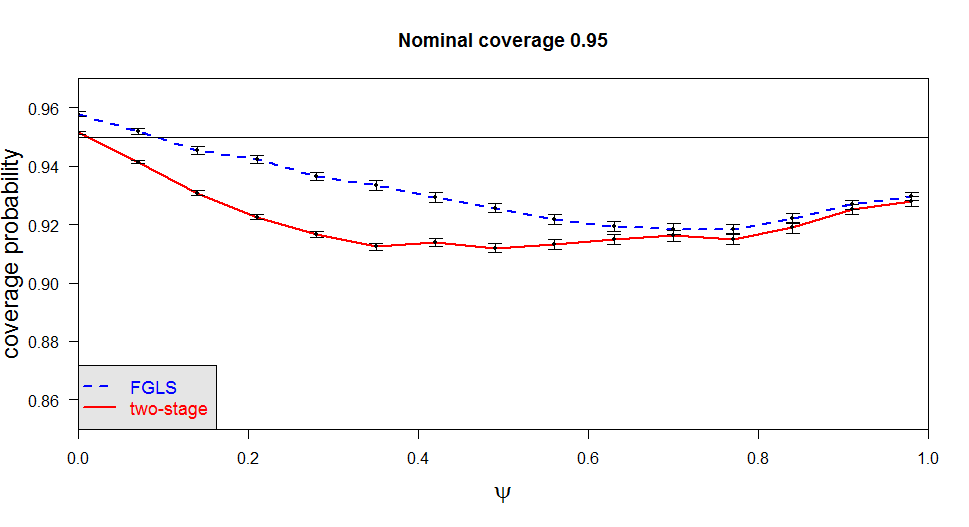}
		\scriptsize
		%		\caption{The design matrix listed in Appendix  \ref{A.3}, where our 
		% parameter of interest is $\beta_5$}
		%		\label{a}
	\end{subfigure}%
	%	\begin{subfigure}{.5\textwidth}
	%		\centering
	%		\includegraphics[width=1\linewidth]{DWandFGLSmat6124.png}
	%		\caption{The design matrix listed in Appendix \ref{A.4}, where our parameter of interest is a linear combination of  $\beta$}
	%		\label{b}
	%	\end{subfigure}
	\\
	\begin{subfigure}{0.5\textheight}
%\begin{subfigure}
		\centering
		\includegraphics[width=1.2\linewidth]{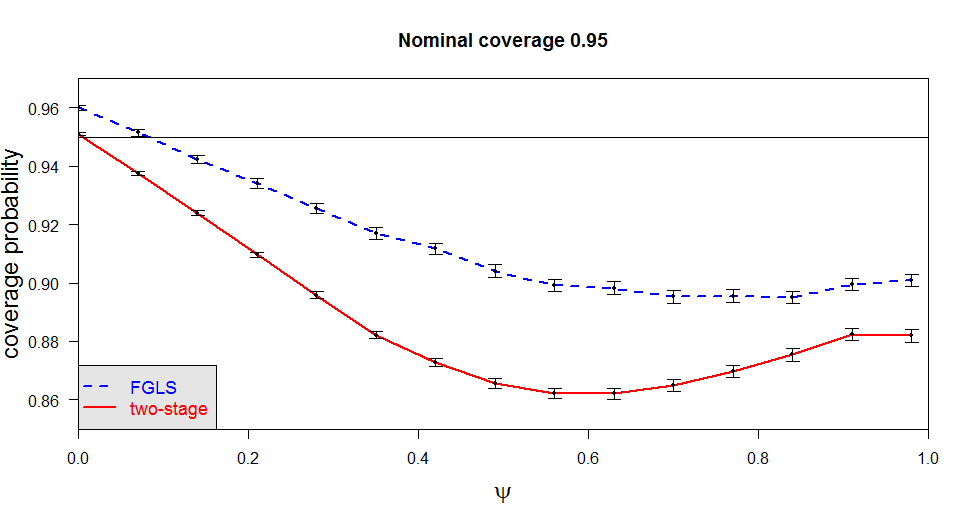}
%\includegraphics[width=1\linewidth]{mat174.png}
		%		\caption{The design matrix listed in Appendix \ref{A.5}, where our 
		% parameter of interest is $\beta_3$}
		%		\label{b}
	\end{subfigure}
	%	\begin{subfigure}{.5\textwidth}
	%		\centering
	%		\includegraphics[width=1\linewidth]{DWandFGLSmat23.png}
	%		\caption{The design matrix listed in Appendix  \ref{A.1}, where our parameter of interest is $\beta_3$}
	%		\label{d}
	%	\end{subfigure}
	\caption{\small The top panel presents graphs of the coverage probability functions for the FGLS and two-stage confidence intervals, each with nominal coverage 0.95, for the chicken demand example. The bottom panel presents graphs of the coverage probability functions for these confidence intervals and the same nominal coverage, for the defense spending example.
	}
	\label{CPDWandFGLS}
\end{figure}

Figure \ref{CPDWandFGLS} presents graphs of the coverage probability functions of the FGLS and two-stage confidence intervals, each with nominal coverage 0.95, for two 
real life data examples. 
The coverage probability for each given value of $\psi \in \{ 0, 0.07, 0.14, \dots, 0.98 \}$ was estimated using 50,000 simulation runs that employ the variance reduction methods described in Section 4 and the Supporting Information. The vertical bars in Figure 1 are approximate 95\% confidence intervals for 
the coverage probabilities
estimated for each value of $\psi$. The computations for this paper were carried out using \texttt{R} programs and packages.

The top panel is for the chicken demand example 
considered on p.333 of Studenmund (2006) and based on the data in Table 6.2 
on p.189. In this example, the response is $Y_t$, the per capita chicken consumption (in pounds) in year $t$, and the model is 
\begin{equation*}
Y_t = \beta_1 + \beta_2 \, PC_t + \beta_3 \, PB_t + \beta_4 \, YD_t + e_t,
\end{equation*}
where $PC_t$ is the price of chicken (in cents per pound), 
$PB_t$ is the price of beef (in cents per pound),
$YD_t$ is the U.S. per capita disposable income (in hundreds of dollars)
and $e_t$ denotes the random error in year $t$. For 
the top panel of 
Figure 1, the parameter of interest
is $\beta_3$. 

The bottom panel is for the defense spending example 
considered on p.342 of Studenmund (2006) and based on the data in Table 9.1 
on pp.343--344. In this example, the linear regression model is
\begin{equation*}
\log(SD_t) = \beta_1 + \beta_2 \, \log(USD_t) + \beta_3 \, \log(SY_t) + \beta_4 \, \log(SP_t) + e_t,
\end{equation*}
where $SDH_t$ is the CIA's ``high'' estimate of Soviet defense expenditure 
%in year $t$ 
(billions of 1970 rubles), $USD_t$ is U.S. defense expenditure 
%in year $t$ 
(billions of 1980 dollars), $SY_t$ is Soviet GNP 
%in year $t$ 
(billions of 1970 rubles), $SP_t$ is the ratio of the number of USSR nuclear warheads ($NR_t$)
to the number of U.S. nuclear warheads ($NU_t$) 
and $e_t$ denotes the random error 
in year $t$.
For 
the bottom panel of
Figure 1, the parameter of interest
is $\beta_4$. 
In both cases, the FGLS confidence interval outperforms the two-stage confidence 
interval, in terms of coverage probability. The resulting recommendation is that 
the FGLS confidence interval should be used, instead of the two-stage confidence 
interval.

%\FloatBarrier

\captionsetup{
	font=normalsize,
	labelfont={bf,sf}}
\captionsetup[sub]{font=scriptsize,labelfont={bf,sf}} 

%\FloatBarrier

In Section 6 we provide two further examples of the comparison of the coverage probabilities of the FGLS and two-stage confidence intervals.
In Section 7 we describe the gains in simulation efficiency achieved using the variance reduction methods described in Section 4 and the Supporting Information in the context of the examples described in Figures 1 and 2.

\section{The OLS and FGLS confidence intervals}

Consider the linear regression model
\begin{equation*}
y = X \beta + e
\end{equation*}
where  $y$ is a  $n$-vector of responses, $X$ is an $n \times p$ known design matrix with linearly independent columns ($n > p$), $\beta$ is an $p$-vector of unknown parameters, $e$ is an $n$-vector of zero-mean random errors. 
We suppose that $\{e_t \}$
is a zero-mean strictly stationary first order autoregressive (AR(1))
process satisfying
\begin{equation*}
e_t = \psi \, e_{t-1} + u_t
\end{equation*}
for all integer $t$, where $\psi$ is an unknown parameter satisfying
$0 \le \psi < 1$, the $u_t$'s are independent and identically normally distributed with zero mean. Let
 $\sigma^2 = E(e_t^2)$, an unknown positive parameter. The restriction to non-negative values
of $\psi$ is very reasonable for many econometric data sets
(see e.g. Wooldridge, 2016, p.378).
Suppose that the parameter of interest is $\theta = a' \, \beta$, where $a$ 
is a specified non-zero $p$-vector. Let our aim be to find a 
confidence interval for $\theta$ with minimum coverage probability $1 - \alpha$.
Henceforth, suppose that the design matrix $X$, $a$ (which is used in the definition of the parameter of interest $\theta$) and 
$1 - \alpha$ are given.

The 
covariance matrix of $e$ is $\sigma^2 \, G(\psi)$, where $ G(\psi)$ is an 
$n \times n$ matrix with $(i,j)$'th element $\psi^{|i - j|}$.
%
% %
% \begin{equation*}
% 	G(\psi) =
% 	\begin{bmatrix}
% 		1          & \psi       & \psi^2     & \psi^3     & \ldots & \psi^{n-1}\\
% 		\psi       & 1          & \psi       & \psi^2     & \cdots & \psi^{n-2}\\
% 		\psi^2     & \psi       & 1          & \psi       & \cdots & \psi^{n-3}\\
% 		\vdots     & \vdots     & \vdots     & \vdots     & \ddots &\vdots\\
% 		\psi^{n-1} & \psi^{n-2} & \psi^{n-2} & \psi^{n-4} & \cdots & 1\\
% 	\end{bmatrix}.
% \end{equation*}
% %
Suppose, for the moment, that $\psi$ is known.
The standard estimator of $\beta$ is 
$\widehat{\beta}(\psi) 
= \big(X^\prime G^{-1}(\psi) X \big)^{-1} X^\prime G^{-1}(\psi) \,  y$.
The resulting estimator of $\theta$ is 
$\widehat{\theta}(\psi)  = a^\prime \widehat{\beta}(\psi)$. Let 
$\widehat{\sigma}^2(\psi)  = \big(y - X \widehat{\beta}(\psi)\big)' \, 
G^{-1}(\psi) \big(y - X \widehat{\beta}(\psi) \big) \big/ m$,
where $m = n - p$.
In other words, $\widehat{\beta}(\psi)$,  $\widehat{\theta}(\psi)$
and $\widehat{\sigma}^2(\psi)$ denote the generalized least squares
estimators of $\beta$, $\theta$ and $\sigma^2$, respectively.
Also let 
$v(\psi)  = Var\big(\widehat{\theta}(\psi)\big) / \sigma^2 = a^\prime \big(X^\prime G^{-1}(\psi) X \big)^{-1}  a$. 
Let $[a \pm b]$ denote the interval $[a - b, a + b]$ ($b>a$).
For $\psi$ known, the standard confidence interval for $\theta$, with coverage $1-\alpha$, is  
\begin{equation*}
J(\psi) =  \left[\widehat{\theta}(\psi)  \pm t_{m , 1-\alpha/2}\left(v(\psi) \right)^{1/2} \;   \widehat{\sigma}(\psi)  \right],
\end{equation*}
where the quantile $t_{m, p}$ is defined as $P(T \leq t_{m, p}) = p$ for $T \sim t_m$. For $\psi = 0$, this confidence interval reduces to the usual
confidence interval for $\theta$, with coverage $1-\alpha$, centred on the 
ordinary least squares (OLS) estimator. We call this the OLS confidence interval.
When $\psi$ is unknown we replace it by the restricted maximum likelihood (REML) estimator $\widehat{\psi}_{\REML}$ in $J(\psi)$ to obtain
the feasible generalized least squares (FGLS) confidence interval $J\big(\widehat{\psi}_{\REML} \big)$. In Appendix A, we describe three estimators of $\psi$, 
including $\widehat{\psi}_{\REML}$. 
Let $e^{\dag} = e / \sigma$, so that $e^{\dag} \sim N(0, G(\psi))$.
The following theorem, proved in Appendix A, is the main result of this section.
\begin{theorem}
	\label{Theorem1}
	% Suppose that the design matrix $X$ is given. 
	The three estimators 
	$\widehat{\psi}$, $\widehat{\psi}_{\ML}$ and $\widehat{\psi}_{\REML}$
	described in Appendix A are all functions of $e^{\dag}$. 
	For both $\widetilde{\psi} = 0$ and $\widetilde{\psi}$ one of these three estimators of $\psi$, the event
	$\big\{\theta \in J \big(\widetilde{\psi} \big)\big\}$
	is equal to the event
	\begin{equation}
	\label{ExpressionForThetaInJ}
	\left\{\big(b(\widetilde{\psi}) \big)^{\prime} e^\dagger 
	\in \Big[0 \pm  
	t_{m,1-\alpha/2} \, \big(v(\widetilde{\psi})\big)^{1/2} \, \big(w(e^\dagger, \widetilde{\psi}) \big)^{1/2} \Big]\right\},
	\end{equation}
	where
	\begin{align*}
	w(e^\dagger, \psi) &= \frac{1}{m} \big(e^\dagger \big)^\prime \,  G^{-1}(\psi) \Big(I - X \, \big(X^\prime G^{-1}(\psi) X \big)^{-1} \, X^\prime \, G^{-1}(\psi)\Big) \, e^\dagger
	\\
	\big(b(\psi) \big)^{\prime} 
	&= a^\prime \big(X^\prime G^{-1}(\psi) X \big)^{-1} \, X^\prime \, G^{-1}(\psi).
	\end{align*}
	Consequently, $P \big(\theta \in J(\widehat{\psi})\big)$,
	$P \big(\theta \in J(\widehat{\psi}_{\ML})\big)$ 
	and 
	$P \big(\theta \in J(\widehat{\psi}_{\REML})\big)$ 
	are functions of $\psi$. 
\end{theorem}

This theorem allows us to easily carry out a numerical comparison of the 
coverage probability functions of the confidence intervals
$J(\widehat{\psi})$,
$J(\widehat{\psi}_{\ML})$ 
and 
$J(\widehat{\psi}_{\REML})$ 
for any given
design matrix $X$, $a$ (which is used in the definition of the parameter of interest $\theta$) and 
$1 - \alpha$ (the desired minimum coverage probability) are given.
These coverage probabilities do \textsl{not} depend on either $\beta$
or $\sigma^2$ and are determined solely by $\psi$. We 
compared these coverage probability functions for the same $X$, $a$  and 
$1 - \alpha$ as those considered in Figure 1. We found that,
in terms of coverage probability, 
$J(\widehat{\psi}_{\REML})$ performs better than $J(\widehat{\psi}_{\ML})$, 
and $J(\widehat{\psi}_{\ML})$ performs better than $J(\widehat{\psi})$.
This finding provides support for the choice we made earlier to always 
construct  
the FGLS confidence interval using the REML estimator of $\psi$. 

 \section{The two-stage confidence interval}

The Durbin-Watson test statistic is  
\begin{equation*}
\widehat{d} = \frac{\sum_{i = 2}^{n} {(r_i - r_{i-1})^2}}{\sum_{i = 1}^{n}{r_i^2}},
\end{equation*}   
where $r_i$ is the $i$'th component of the vector $r =  \big(I - X (X^\prime X)^{-1} X^\prime \big) \, e$ of residuals from the 
model fitted by OLS. It may be shown that 
\begin{equation*}
\widehat{d} = \frac{r^\prime \, B \, r}{r^\prime \, r},
\end{equation*}
where
\begin{equation*}
B=
\begin{bmatrix}
1      & -1     & 0      & \cdots & \cdots & 0      \\
-1     & 2      & \ddots & \ddots &        & \vdots \\
0      & \ddots & \ddots & \ddots & \ddots & \vdots \\
\vdots & \ddots & \ddots & \ddots & \ddots & 0      \\
\vdots &        &\ddots  & \ddots & 2      & -1     \\
0      & \cdots & \cdots & 0      & -1     & 1      \\
\end{bmatrix}.
\end{equation*}
Dividing the numerator and denominator of this expression for $\widehat{d}$
by $\sigma^2$, we find that the Durbin-Watson test statistic
\begin{equation}
\label{DurbinWatson3}
\widehat{d} 
%   = \frac{(e^\dagger)^\prime \, \big(I - X (X^\prime X)^{-1} X^\prime \big)^\prime 
%   		\;  B  \;  
%   		\big(I - X (X^\prime X)^{-1} X^\prime \big) \, e^\dagger}
%   	{(e^\dagger)^\prime \, \big(I - X (X^\prime X)^{-1} X^\prime \big) \, e^\dagger}.
= \frac{\big(r^{\dag}\big)^{\prime} \, B \, r^{\dag}}{\big(r^{\dag}\big)^{\prime} \, r^{\dag}},
\end{equation}
where
$r^{\dag} = r / \sigma$, so that $r^{\dag} = \big(I - X (X^\prime X)^{-1} X^\prime \big) \, e^{\dag}$.
We use this test statistic as follows
to test the null hypothesis that $\psi =0$ against the alternative hypothesis that $\psi > 0$. If $\widehat{d} > c(\widetilde{\alpha})$ then we accept this null hypothesis;
otherwise we reject this null hypothesis. Here 
$c(\widetilde{\alpha})$ is defined to be the value of $c$ such that 
under the null hypothesis 
$P(\widehat{d} \le c) = \widetilde{\alpha}$, a specified 
test size. The method used to compute $c(\widetilde{\alpha})$ is described in 
Appendix B. Henceforth, suppose that $\widetilde{\alpha}$ is given.

Consider the following two-stage procedure. We carry out a Durbin-Watson pretest of the null hypothesis that $\psi =0$ against the alternative hypothesis that $\psi > 0$. 
If this null hypothesis is accepted then we use the OLS confidence interval, with nominal coverage $1 - \alpha$; otherwise we use an FGLS confidence interval, with nominal coverage $1 - \alpha$. We call this the two-stage 
confidence interval, with desired minimum coverage $1 - \alpha$, and we denote it by $K$.
In other words, 
\begin{equation*}
K = 
\begin{cases}
J(0) &\text{if} \ \ \widehat{d} > c(\widetilde{\alpha})\\
J(\widehat{\psi}_{\REML})  &\text{otherwise}.
\end{cases}
\end{equation*}

The following theorem, which is the main result of this section, 
is proved in Appendix B using Theorem \ref{Theorem1} and the 
expression \eqref{DurbinWatson3} for the Durbin-Watson statistic 
$\widehat{d}$. 
\begin{theorem}
	\label{Theorem2}
	The coverage probability of the two-stage confidence interval 
	$K$,
	$P(\theta \in K)$, is a function of $\psi$.
\end{theorem}
Using this theorem, we can easily carry out a numerical comparison of the 
coverage probability functions of the FGLS confidence interval $J(\widehat{\psi}_{\REML})$ and the two-stage confidence interval $K$, 
for the 
same values of $X$, $a$  and 
$1 - \alpha$.
These coverage probabilities do \textsl{not} depend on either $\beta$
or $\sigma^2$ and are determined solely by $\psi$.
As described in the two next sections,
we compute these coverage probability functions by simulation.

 \section{Computation of the coverage probability of the FGLS 
	confidence interval by simulation}

We may compute the coverage probability of the FGLS confidence interval $J(\widehat{\psi}_{\REML})$ using ``brute force'' simulation 
as follows. Suppose that we carry out $M$ independent simulation runs.
On the $k$'th simulation run we compute an observation of $e^{\dag}$
and then record whether or not the event \eqref{ExpressionForThetaInJ} occurs. 
The total number of occurrences of this event has a Binomial$(M, p)$
distribution, where $p = P(\theta \in J(\widehat{\psi}_{\REML}))$.
The estimator of $p$ and its standard error are found using the well-known
properties of this distribution.

However, a better way to compute this coverage probability by simulation
is to use variance reduction as follows. 
Let
\begin{equation*}
\boldsymbol{1}({\cal B}) =
\begin{cases}
1 &\text{if} \   \   \ \mathcal{B} \  \text{is true}
\\
0      &\text{if}   \   \ \mathcal{B}  \  \text{is false},
\end{cases}
\end{equation*}
where $\mathcal{B}$ is an arbitrary statement. Also let
\begin{equation}
\label{Defn_h}
h(e^\dagger,\psi )
= \boldsymbol{1} \left(\big(b(\psi) \big)^{\prime} e^\dagger 
\in \Big[0 \pm  
t_{m,1-\alpha/2} \, \big(v(\psi)\big)^{1/2} \, \big(w(e^\dagger, \psi) \big)^{1/2} \Big]\right),
\end{equation}
so that, by Theorem \ref{Theorem1}, 
$P \big(\theta \in J(\widehat{\psi}_{\REML})\big) 
= E \big(h(e^\dagger,\widehat{\psi}_{\REML}) \big)$.
We expect that, with probability close to 1, 
$h(e^\dagger,\widehat{\psi}_{\REML})$ will be close to
$h(e^\dagger,\psi)$, particularly for large $n$.
Note that $E \big(h(e^\dagger,\psi) \big) = 1 - \alpha$.
This motivates our use of $h(e^\dagger,\psi)$ as a control variate.
Therefore 
\begin{equation*}
P \big(\theta \in J(\widehat{\psi}_{\REML})\big)
= 1 - \alpha 
+ E\Big(h\big(e^\dagger,\widehat{\psi}_{\REML}\big) 
- h \big(e^\dagger,\psi \big) \Big).
\end{equation*}
We expect that $Var \big(h\big(e^\dagger,\widehat{\psi}_{\REML}\big) 
- h \big(e^\dagger,\psi \big) \big)$ will be much less
than $Var \big(h\big(e^\dagger,\widehat{\psi}_{\REML}\big) 
\big)$, particularly for large $n$. The resulting computation of 
$P\big(\theta \in J(\widehat{\psi}_{\REML})\big)$ by simulation is as follows.
Suppose that we carry out $M$ independent simulation runs.
On the $k$'th simulation run we compute an observation of $e^{\dag}$
and then record the value of 
$h\big(e^\dagger,\widehat{\psi}_{\REML}\big) 
- h \big(e^\dagger,\psi \big)$.
Then $1 - \alpha + \text{(sample mean of these values)}$ is the estimate
of $P\big(\theta \in J(\widehat{\psi}_{\REML})\big)$, with standard error
$\big((\text{sample variance of these values}) \big/ M \big)^{1/2}$.

\section{Computation of the coverage probability of the two-stage
	confidence interval by simulation}

We may compute the coverage probability of the two-stage confidence interval $K$ using ``brute force'' simulation 
as follows. Suppose that we carry out $M$ independent simulation runs.
On the $k$'th simulation run we compute an observation of $e^{\dag}$
and then record if the event $\{\theta \in K\}$ occurs. 
The total number of occurrences of this event has a Binomial$(M, p)$
distribution, where $p = P(\theta \in K)$.
The estimator of $p$ and its standard error are found using the well-known
properties of this distribution.
However, a better way to compute this coverage probability by simulation
is to use the variance reduction method described in the Supporting Information.

 \section{Two more examples of the comparison of the coverage probabilities
	of the FGLS and two-stage confidence intervals}

\begin{figure}[]
	\begin{subfigure}{0.5\textheight}
		\centering
		\includegraphics[width=1.2\linewidth]{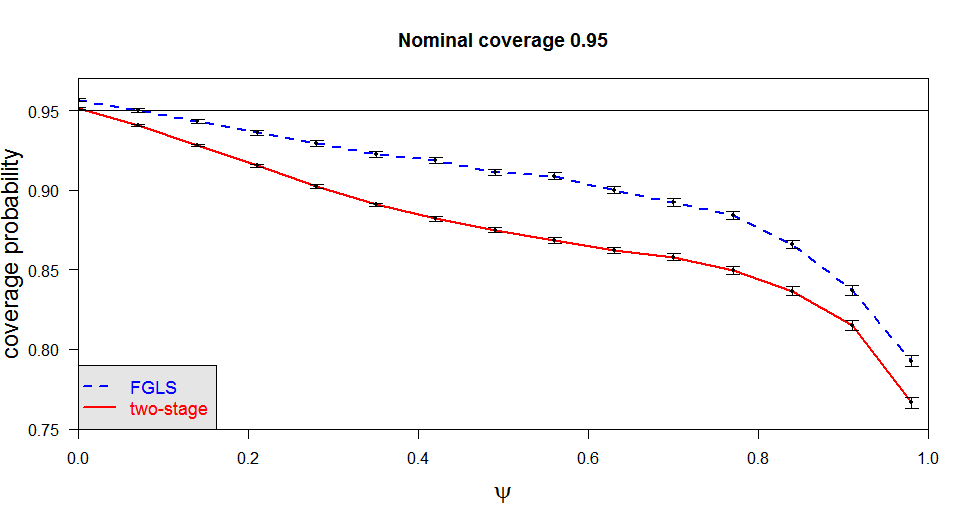}
		\scriptsize
	\end{subfigure}%
	\\
	\begin{subfigure}{0.5\textheight}
		\centering
		\includegraphics[width=1.2\linewidth]{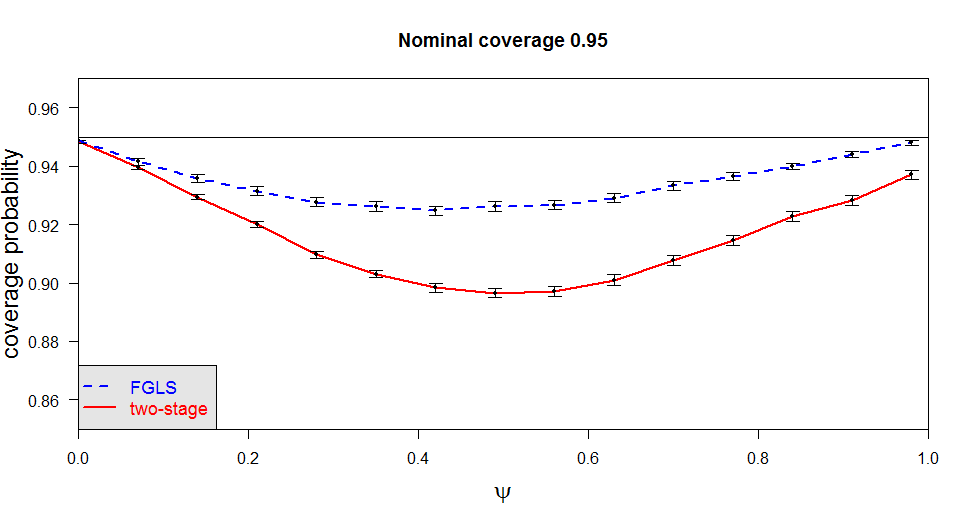}
	\end{subfigure}
	\caption{\small The top panel presents graphs of the coverage probability functions for the FGLS and two-stage confidence intervals, each with nominal coverage 0.95, for the fish demand  example. The bottom panel presents graphs of the coverage probability functions for these confidence intervals, and the same nominal coverage, for the ice cream consumption example.
	}
	\label{CPDWandFGLS1}
\end{figure}

Figure \ref{CPDWandFGLS1} presents graphs of the coverage probability functions of the FGLS and two-stage confidence intervals for two real life data examples. Both of these confidence intervals have nominal coverage 0.95.
The coverage probability for each given value of $\psi \in \{ 0, 0.07, 0.14, \dots, 0.98 \}$ was estimated using 50,000 simulation runs that employ the variance reduction methods described in Section 4 and the Supporting Information. The vertical bars in Figure \ref{CPDWandFGLS1} are approximate 95\% confidence intervals for 
the coverage probabilities
estimated for each value of $\psi$.

The top panel of Figure \ref{CPDWandFGLS1} is for the fish demand example 
considered on p.334 Studenmund (1992) and based on the data in Table 8.1 
on p.290, which was obtained from Historical Statistics of the US, Colonial Times to 1970 part 1. In this example, the linear regression model is
\begin{equation*}
F_t = \beta_1 + \beta_2 \, RP_t + \beta_3 \, \log(Yd_t) + \beta_4 \, D_t + e_t,
\end{equation*}
where $F_t$ is the average pounds of fish consumed per capita in year $t$, 
$RP_t$ is the price of fish relative to beef in year $t$, 
$Yd_t$ is the real per capita disposable income in year $t$ (in billions of dollars), 
$D_t$ is a dummy variable equal to zero in years before 1966 and one afterwards  
and $e_t$ denotes the random error 
in year $t$.
For this panel of Figure 2, the parameter of interest is  $\beta_3$.
This panel provides an illustration of the case that, while the FGLS confidence interval outperforms the
two-stage confidence interval in terms of the coverage probability function, 
the coverage probability performance of both of these intervals drops 
substantially as $\psi$ approaches 1.

The bottom panel of Figure \ref{CPDWandFGLS1} is for consumption of ice cream  example considered on p.104  Verbeek (2004). This data is listed by Hildreth and Lu (1960) and consists of 30 four-weekly observations from 18 March 1951 to 11 July 1953. In this example, the response is $y_t$, the consumption of ice cream per head (pints)  at time $t$ (measured in consecutive four-weekly segments), and the model is 
\begin{equation*}
y_t = \beta_1 + \beta_2 \, X_{t1} + \beta_3 \, X_{t2}  + \beta_4 \, X_{t3} + e_t,
\end{equation*}
where $X_{t1}$ is the average family income per week (in US Dollars),
$X_{t2}$ is the price of ice cream (per pint),
$X_{t3}$ is the average temperature (in Fahrenheit)
and $e_t$ denotes the random error at time $t$. For this panel of Figure \ref{CPDWandFGLS1}, the parameter of interest is  $\beta_3$. 
This panel provides yet another illustration of the case that the 
FGLS confidence interval outperforms the
two-stage confidence interval in terms of the coverage probability function.
This panel also provides an illustration of the 
case that both the FGLS and two-stage confidence intervals have coverage probability very close to the nominal coverage 0.95 for $\psi = 0$.

\captionsetup{
	font=normalsize,
	labelfont={bf,sf}}
\captionsetup[sub]{font=scriptsize,labelfont={bf,sf}} 

\section{Efficiency of the simulation methods that use variance reduction}

All of the computations reported in this paper were carried out on a PC with Intel i7 CPU and 32Gb of RAM. 
The coverage probabilities of the FGLS and two-stage confidence 
intervals, each with nominal coverage 0.95, were computed using 50,000 simulation runs,
with the variance reductions described in Section 4 and the Supporting Information, 
for each $\psi \in \{ 0, 0.07, 0.14, \dots, 0.98 \}$ and
for both panels of both figures. 
The use of variance reduction has greatly increased the efficiency of the computations of the graphs
shown in both Figures \ref{CPDWandFGLS} and \ref{CPDWandFGLS1}.

To assess the improvement in the efficiency of these computations, we also
computed the coverage probabilities of the FGLS and two-stage confidence 
intervals, each with nominal coverage 0.95, using 50,000 simulation runs,
without variance reduction, 
for each $\psi \in \{ 0, 0.07, 0.14, \dots, 0.98 \}$ and
for both panels of both figures. Using the standard measure of 
relative efficiency given e.g. on p.51 of Hammersley and Handscomb (1964),
we then found the computation time required to estimate these coverage probabilities 
with the same accuracy
(i.e. with the same standard error)
as when the variance reduction methods described in Section 4 and the Supporting Information were employed.
The results of these computations are presented in 
Tables 1 and 2 below, which concern the FGLS and two-stage confidence 
intervals, respectively.
In Table 1 variance reduction leads to the computer times being reduced by factors ranging between 1.75 to 2.82. 
In Table 2  variance reduction leads to the computer times being reduced by factors ranging between 2.15 to 3.48.

\begin{table}[h]
	\renewcommand\thetable{1}
	\centering
	\caption{Computation times for simulation estimates, with the same accuracy, of the coverage probabilities of the FGLS confidence interval for 
		each $\psi \in \{ 0, 0.07, 0.14, \dots, 0.98 \}$, with and without the variance reduction described in Section 4.}
	\begin{tabular}{|c|c|c|}
		\hline
		& With variance reduction & Without variance reduction   \\
		%\hline
		\hline
		Fig.1 top panel \ \ \ \ \   &   82 mins       &   208 mins        \\ 
		\hline
		Fig.1 bottom panel &  51 mins        &  102 mins         \\
		\hline
		Fig.2 top panel  \ \ \ \ \     &    103 mins      &   180 mins        \\
		\hline
		Fig.2 bottom panel &    115 mins      &   324 mins        \\
		\hline
	\end{tabular}
	\label{table1}
\end{table}

\begin{table}[h] 
	\renewcommand\thetable{2}
	\centering
	\caption{Computation times for simulation estimates, with the same accuracy,  of the coverage probabilities of the two-stage confidence interval
		for 
		each $\psi \in \{ 0, 0.07, 0.14, \dots, 0.98 \}$,
		with and without the variance reduction described in Section 5.}
	\begin{tabular}{|c|c|c|}
		\hline
		& With variance reduction & Without variance reduction   \\
		%\hline
		\hline
		Fig.1 top panel \ \ \ \ \      &   81 mins       &    282 mins       \\ 
		\hline
		Fig.1 bottom panel  &    53 mins      &    134 mins       \\
		\hline
		Fig.2 top panel \ \ \ \ \      &    118 mins      &    268 mins       \\
		\hline
		Fig.2 bottom panel  &   118 mins        &    254 mins        \\
		\hline
	\end{tabular}
	\label{table2}
\end{table} 

\section{Remarks}

\textsl{Remark 8.1.} The \texttt{R} programs used to compute the coverage probabilities of the FGLS 
		and two-stage confidence intervals, using simulation with the variance reduction methods described in Section 4 and the Supporting Information, were checked for correctness
		in the following two ways, for all of the examples considered in the paper.
		Firstly, these coverage probabilities were computed by simulation, without 
		variance reduction, using simulations of $e^{\dag}$
		and the expression
		\eqref{ExpressionForThetaInJ}
		with $\widetilde{\psi}$ replaced by 
		$\widehat{\psi}_{\REML}$ and 0, respectively, as appropriate. 
		Secondly, these coverage probabilities were also computed by simulation, 
		without variance reduction, using simulations of 
		$y = X \beta + e$ for the particular case that $\beta = 0$ (which implies that
		$\theta = 0$) and $\sigma^2 = 1$.

\bigskip

\noindent \textsl{Remark 8.2.} 
	Straightforward analogues of Theorem \ref{Theorem2} and the variance reduction methods described in the Supporting Information hold if we replace the Durbin-Watson test statistic
		by the so-called ``t-statistic''
		\begin{equation*}
		\dfrac{\widehat{\psi}}
		{\left(\frac{1}{n-2} \sum_{t=2}^n \big(r_t -\widehat{\psi} \, r_{t-1} \big)^2 \big/ \sum_{s=1}^{n-1} r_s^2\right)^{1/2}},
		\end{equation*}
		where $\widehat{\psi}$ is the estimator of $\psi$ described in Appendix A.

\bigskip

\noindent \textsl{Remark 8.3.} Theorems 2.1 and 3.2 and the variance reduction method described in Section 4 extend in the obvious way to the case that $\{e_t \}$
is assumed to be an an ARMA$(\ell_1, \ell_2)$ process for any given $\ell_1$
and $\ell_2$ ($\ell_1 + \ell_2 >0$).

\bigskip

\noindent \textsl{Remark 8.4.} The framework that we use for the construction of confidence intervals for $\theta$ does not preclude the consideration of misspecification of the model for the autocorrelations of the random errors. We expect that for moderate levels of misspecification and moderate sample sizes, the result will be a negligible change in the performance of the confidence interval for $\theta$ constructed assuming that there is no misspecification. Such an assertion can easily be checked using a sensitivity analysis in which the actual data generating process for the random errors is not included in the assumed family of parametric models.

\bigskip

\noindent \textsl{Remark 8.5.} 
An alternative framework for the construction of confidence intervals for $\theta$ is 
to use 
a confidence interval centred on the OLS estimator, but with the correct standard error estimated using a heteroskedasticity and autocorrelation consistent (HAC) estimator (see e.g. Andrews, 1991). A remarkable feature of this estimator is that it is consistent for virtually arbitrary autocorrelations. 
One should, however, not
lose sight of the fact that the OLS estimator is typically inefficient by comparison with competitors of the type described by Wooldridge (2016, p.390).
Also, 
confidence intervals based on this estimator and standard error estimated by HAC can perform poorly, in terms of coverage
probability, for moderate values of $n$.
In this alternative framework there seems to be little motivation
for carrying out any preliminary hypothesis test.

\section{Discussion}

It is common in applied econometrics to carry out preliminary data-based 
model selection, using preliminary hypothesis tests or minimizing a criterion such as the Akaike Information Criterion. This is frequently followed by the construction of a confidence interval for a scalar parameter of interest, using the same data, based on the assumption that the selected model had been given to us \textit{a priori}, as the true model. This assumption is false  because (a) the preliminary model selection sometimes chooses the wrong model and (b) the data used to choose the model is re-used for
the construction of the confidence interval without due acknowledgement.
It is important to delineate those models and model selection procedures for which the post-model-selection confidence interval has poor coverage properties. A review of some of the literature that carries out 
this delineation is provided by Kabaila (2009).
Kabaila, Mainzer and Farchione (2015, 2017) and the present paper 
extend this delineation project to model selection procedures that are 
of particular interest in the field of econometrics.

In the present paper we consider
a preliminary data-based
selection of a time series model for the random errors in a linear regression model. We provide the tools needed to assess the effect of this
preliminary model selection on  the coverage probability of
a confidence interval for a given linear combination of the regression parameters and a given design matrix. 
The first tool is to show that the coverage probabilities of both the 
FGLS and two-stage confidence intervals do not depend on either the regression
parameter vector or the variance of the random error. The second tool is to 
provide methods of variance reduction for the simulations used to estimate these coverage probability functions, leading to the provision of these estimates
in a reasonable amount of time.
Our proposal is that this assessment be carried out 
on a case-by-case basis for each given design matrix and parameter of interest. 
Since this assessment is carried out prior to the examination of the observed
response vector, it can validly be used to decide whether the FGLS confidence interval or the two-stage confidence interval should be used.

    % \bigskip

    %\newpage

\section*{Appendix A: FGLS confidence intervals}
\renewcommand{\theequation}{A.\arabic{equation}}
\renewcommand{\thesection}{A}
\setcounter{equation}{0}
  
    \medskip
    
    \noindent \textbf{The three estimators of $\boldsymbol{\psi}$ considered}
    
    \medskip
    
    We consider three estimators of $\psi$. The first of these is 
    \begin{equation*}
    \widehat{\psi} = \frac{\sum_{t=2}^n r_t \, r_{t-1}}{\sum_{t=1}^n r_t^2},
    \end{equation*}
    where $r_t$ is the $t$'th component of the vector $r =  \big(I - X (X^\prime X)^{-1} X^\prime \big) \, e$ of residuals from the 
    model fitted by OLS. This estimator is the 
    sample first order autocorrelation of the residuals from this fitted model. 
    
    The second estimator is $\widehat{\psi}_{\ML}$,  the maximum likelihood estimator of $\psi$, is obtained (see e.g. Cooper and Thompson, 1977) by maximizing
    \begin{equation}
    \label{MLloss}
    -\frac{n}{2} \log 
    \left(\frac{S \big(\widehat{\beta}(\psi), \psi \big)}{n}\right)
    - \frac{1}{2} \log \left(|G(\psi)|\right)
    \end{equation}
    with respect to $\psi \in [0, 1)$,
    where $S(\beta,\psi) = (y - X \beta)^\prime \, G^{-1}(\psi) \, (y - X \beta)$.
    %%     
    %\begin{equation*}
    %S(\beta,\psi) = (y - X \beta)^\prime \, G^{-1}(\psi) \, (y - X \beta).
    %\end{equation*}
    %%%
    %and
    %%
    %\begin{equation*}
    %\widehat{\beta} 
    %= \big(X^\prime G^{-1}(\psi) X \big)^{-1} \, X^\prime G^{-1}(\psi) \,  y.
    %\end{equation*}
    %%  
    The third estimator is $\widehat{\psi}_{\REML}$, the restricted maximum
    likelihood estimator of $\psi$, is obtained 
    (see e.g. Cheang and Reinsel, 2000) by maximizing
    \begin{equation}
    \label{REMLloss}
    - \frac{m}{2}\log 
    \left(\frac{S\big(\widehat{\beta}(\psi) ,\psi\big)}{m} \right) - \frac{1}{2}\log(|G(\psi)|) - \frac{1}{2}\log \left(|X^\prime G^{-1}(\psi) X|\right)
    \end{equation}
    with respect to $\psi \in [0, 1)$.
    
    \medskip
    
    \noindent \textbf{Proof of Theorem \ref{Theorem1}}
    
    \medskip
    
    Let $e^{\dag} = e / \sigma$ and note that $e^{\dag} \sim N(0, G(\psi))$. We show that $\widehat{\psi}$, $\widehat{\psi}_{\ML}$ and $\widehat{\psi}_{\REML}$ are all
    functions of $e^{\dag}$.
    Let
    $r^{\dag} = r / \sigma$, so that $r^{\dag} = \big(I - X (X^\prime X)^{-1} X^\prime \big) \, e^{\dag}$. Division of the numerator and denominator of the expression for $\widehat{\psi}$ by $\sigma^2$ shows that 
    \begin{equation*}
    \widehat{\psi} = \frac{\sum_{t=2}^n r^{\dag}_t \, r^{\dag}_{t-1}}{\sum_{t=1}^n \big(r^{\dag}_t \big)^2}.
    \end{equation*}
    In other words, $\widehat{\psi}$ is a function of $e^{\dag}$.
    
    The proofs that $\widehat{\psi}_{\ML}$ and $\widehat{\psi}_{\REML}$ are 
    functions of $e^{\dag}$ are almost identical and so we present only the 
    proof for $\widehat{\psi}_{\ML}$. It follows from
    \begin{equation*}
    y - X \, \widehat{\beta}(\psi)
    = \Big(I - X \big(X^\prime G^{-1}(\psi) X \big)^{-1} X^\prime G^{-1}(\psi) \Big) \, e
    \end{equation*}
    that
    \begin{align*}
    %\label{ExpressionSforSimulations}
    S \big(\widehat{\beta}(\psi),\psi \big)
    &=  (y - X \widehat{\beta}(\psi))^\prime \, G^{-1}(\psi) 
    \, (y - X \widehat{\beta}(\psi))
    \\
    &= e^\prime \, G^{-1}(\psi)\Big(I - X \big(X^\prime G^{-1}(\psi) X \big)^{-1} X^\prime G^{-1}(\psi)\Big) \, e.
    \end{align*}
    Thus 
    \begin{equation*}
    %\label{ExpressionSforSimulations}
    \frac{S \big(\widehat{\beta}(\psi),\psi \big)}{\sigma^2} 
    = \big(e^{\dag} \big)^\prime \, G^{-1}(\psi)\Big(I - X \big(X^\prime G^{-1}(\psi) X \big)^{-1} X^\prime G^{-1}(\psi)\Big) \, e^{\dag},
    \end{equation*}
    where $\sigma^2$ denotes the true parameter value. Now the criterion
    \eqref{MLloss} is equal to 
    \begin{equation*}
    %\label{MLcriterionModified}
    -\frac{n}{2} 
    \log \left(\frac{S\big(\widehat{\beta}(\psi), \psi \big)}{n \, \sigma^2}\right)
    -\frac{n}{2} 
    \log \left(\sigma^2\right)
    - \frac{1}{2} \log \left(|G(\psi)|\right),
    \end{equation*}
    where $\sigma^2$ denotes the true parameter value. Thus maximizing 
    \eqref{MLloss}
    with respect to $\psi \in [0, 1)$ is equivalent to maximizing
    \begin{equation*}
    %\label{MLcriterionModified}
    -\frac{n}{2} 
    \log \left(\frac{S\big(\widehat{\beta}(\psi), \psi \big)}{n \, \sigma^2}\right)
    - \frac{1}{2} \log \left(|G(\psi)|\right),
    \end{equation*}
    with respect to $\psi \in [0, 1)$, where $\sigma^2$ denotes the true parameter value.
    Consequently, 
    $\widehat{\psi}_{\ML}$ is a function of $e^{\dag}$.

    Suppose that 
    $\widetilde{\psi}$ is one of the three estimators $\widehat{\psi}$,
    $\widehat{\psi}_{\ML}$ and
    $\widehat{\psi}_{\REML}$
     of $\psi$. Also suppose that
    the nominal coverage, $1 - \alpha$, of the confidence interval $J \big(\widetilde{\psi} \big)$
    is given. Then
    \begin{align*}
    \big\{\theta \in J \big(\widetilde{\psi} \big)\big\}
    &= \left\{\theta \in \left[\widehat{\theta}(\widetilde{\psi} )  \pm t_{m , 1-\alpha/2}\left(v(\widetilde{\psi} ) \right)^{1/2} \   \widehat{\sigma}(\widetilde{\psi} )  \right]
    \right\}
    \\
    &= \left\{\frac{\widehat{\theta}(\widetilde{\psi} ) - \theta}{\sigma} \in \left[0 \pm t_{m , 1-\alpha/2}\left(v(\widetilde{\psi} ) \right)^{1/2} \   \frac{\widehat{\sigma}(\widetilde{\psi} )}{\sigma}  \right]
    \right\}
    \\
    &= \left\{a^{\prime} \left(\frac{\widehat{\beta}(\widetilde{\psi} ) - \beta}{\sigma} \right)
    \in \left[0 \pm t_{m , 1-\alpha/2}\left(v(\widetilde{\psi} ) \right)^{1/2} \   \frac{\widehat{\sigma}(\widetilde{\psi} )}{\sigma}  \right]
    \right\}.
    \end{align*}
    Since $\widehat{\sigma}^2(\psi) =S \big(\widehat{\beta}(\psi),\psi \big) \big/ m$,
    \begin{align*}
    \widehat{\sigma}^2(\psi)
    % &= \frac{1}{m} \,\big(y - X \widehat{\beta}(\psi)\big)' \, 
    % G^{-1}(\psi) \big(y - X \widehat{\beta}(\psi) \big)
    % \\
    &= \frac{1}{m} \, e^\prime \, G^{-1}(\psi) \Big(I - X \big(X^\prime G^{-1}(\psi) X\big)^{-1} X^\prime G^{-1}(\psi) \Big) e.
    \end{align*}
    Therefore,
    \begin{equation*}
    \widehat{\sigma}^2(\widetilde{\psi})= \frac{1}{m} \, e^\prime \, G^{-1}(\widetilde{\psi}) \Big(I - X \big(X^\prime G^{-1}(\widetilde{\psi}) X \big)^{-1} X^\prime G^{-1}(\widetilde{\psi}) \Big) e
    \end{equation*}
    Hence    
    \begin{align*}
    \frac{ \widehat{\sigma}^2(\widetilde{\psi})}{\sigma^2} 
    &= \frac{1}{m} (e^\dagger)^\prime \, G^{-1} \, \big(\widetilde{\psi}) \, 
    \Big(I - X \big(X^\prime G^{-1} (\widetilde{\psi}) X \big)^{-1} \,
    X^\prime \, G^{-1} (\widetilde{\psi})\Big) \, e^\dagger\
    \\
    &= w(e^\dagger, \widetilde{\psi}).
    \end{align*}
    Also note that
    \begin{equation*}
    a^{\prime} \left(\frac{\widehat{\beta}(\widetilde{\psi} ) - \beta}{\sigma} \right) = (b(\widetilde{\psi}))^{\prime} \, e^\dagger.
    \end{equation*}
    Thus the event $\big\{\theta \in J \big(\widetilde{\psi} \big)\big\}$ is equal to 
    \eqref{ExpressionForThetaInJ}.
    
    \hfill $\qed$
    
    \section*{Appendix B: The two-stage confidence interval}
    \renewcommand{\theequation}{B.\arabic{equation}}
    \renewcommand{\thesection}{B}
    \setcounter{equation}{0}
    
    \medskip
    
    \noindent \textbf{Computation of $\boldsymbol{P \big(\widehat{d} > c \big)}$
    		%for $\boldsymbol{\psi = 0}$
    }
    
    \medskip
    
    The cutoff  $c(\widetilde{\alpha})$ is defined to be the value of $c$ such that, 
    under the null hypothesis $\psi = 0$, 
    $P(\widehat{d} \le c) = \widetilde{\alpha}$, a specified 
    test size. 
    Observe that
    \begin{align*}
    P \big(\widehat{d }> c \big) 
    & = P \left( \frac{\big(r^{\dag}\big)^{\prime} \, B \, r^{\dag}}{\big(r^{\dag}\big)^{\prime} \, r^{\dag}} > c \right)
    \\
    & = P\Big( \big(r^{\dag}\big)^{\prime} 
    \, ( B  -    c I  ) \, r^{\dag} > 0 \Big) 
    \\
    &= P\Big((e^\dagger)^\prime \  \big(I - X (X^\prime X)^{-1} X^\prime \big) \, ( B  -  cI\ ) \,
    \big(I - X (X^\prime X)^{-1} X^\prime \big) \  e^\dagger > 0 \Big)
    \end{align*}
    where $ e^\dagger \sim N(0,  G(\psi))$.
    We compute the $P \big(\widehat{d} > c \big)$ using the method of 
    Imhof (1961). This is done using the \texttt{Imhof} function in the \texttt{CompQuadForm}  package in \texttt{R}. We compute 
    $\big(X^\prime X \big)^{-1}$ using the QR decomposition of $X$.
    
    \medskip
    
    \noindent \textbf{Proof of Theorem \ref{Theorem2}}
    
    \medskip
    
    Observe that
    \begin{align*}
    \big\{\theta \in K \big\} 
    &= \Big(\big\{\theta \in K \big\} \cap \big\{H_0 \ \text{accepted} \big\} \Big)
    \cup \Big(\big\{\theta \in K \big \} \cap \big \{H_0 \ \text{rejected} \big\} \Big)
    \\
    &= \Big(\big\{\theta \in J(0) \big\} \cap 
    \big\{\widehat{d} > c(\widetilde{\alpha})\big\} \Big)
    \cup \Big(\big\{\theta \in J(\widehat{\psi}_{\REML}) \big \} \cap \big \{\widehat{d} \le c(\widetilde{\alpha}) \big\} \Big).
    \end{align*}
    Now $\big\{\theta \in J(0)\big\}$
    is equal to \eqref{ExpressionForThetaInJ} with $\widetilde{\psi} = 0$, 
    $\big\{\theta \in J \big(\widehat{\psi}_{\REML} \big)\big\}$ is equal to 
    \eqref{ExpressionForThetaInJ} with $\widetilde{\psi}$ replaced by
    $\widehat{\psi}_{\REML}$,
    $\widehat{\psi}_{\REML}$ is a function of $e^{\dag}$ (by Theorem \ref{Theorem1})
    and the Durbin-Watson test statistic $\widehat{d}$ satisfies     \eqref{DurbinWatson3}. Hence whether or not the event $\{\theta \in K\}$ 
    occurs is determined by the random vector $e^{\dag}$, which has an $N(0, G(\psi))$ distribution.
    Thus
    $P(\theta \in K)$ is a function of $\psi$.

    \hfill $\qed$
    
    \noindent \textbf{References}
    
    \medskip

    %  \rf Abrahamse A.P.J., Koerts J. (1969).  A comparison between the power of the Durbin-Watson test and the power of the BLUS test. \textit{Journal of the American Statistical Association}, 64, pp 938--948. 
    %
    %\smallskip
    %
    %  \rf Anderson, T.W. (1971). The Statistical Analysis of Time Series. Wiley, New York .
    %  
    % \smallskip

    \rf Andrews, D.W.K. (1991), Heteroskedasticity and autocorrelation consistent covariance matrix estimation. \textit{Econometrica},
    59, 817--858.
    
    \rf Anselin, L. (2006), Spatial econometrics. Pages 
    931--932 of Palgrave Handbook of Econometrics:  Vol 1, Econometric Theory.
    (T.C. Mills and K. Patterson eds.).
    Palgrave Macmillan, Basingstoke.
    
    \smallskip
    
    \rf  Berthouex, P.M. and  L.C. Brown (2002), Statistics for environmental engineers, 2nd edition, CRC, Boca Raton, FL.

    \smallskip
    
    \rf Cheang, W-K. and G.C. Reinsel (2000), Bias reduction of autoregressive estimates
    in time series regression model through restricted maximum likelihood. \textit{Journal of the American Statistical Association}, 95, 1173--1184.

    \smallskip
    
    \rf Cooper, D.M. and R. Thompson (1977), A note on the estimation of the 
    parameters of the autoregressive-moving average process. 
    \textit{Biometrika}, 64, 625--628.

    \smallskip
    
    \rf Folmer, H. (1988), Autocorrelation pre-testing in linear models with AR(1) errors. Pages 39--55 of On Model Uncertainty and its Statistical Implications,
    Proceedings of a Workshop, Held in Groningen, The Netherlands, September 25--26, 1986 (Theo K. Dijkstra ed.). Springer-Verlag, Berlin. 
    
    \smallskip

    \rf Giles, J.A. and D.E.A. Giles (1993), Pre-test estimation and testing in econometrics: recent developments. \textit{Journal of Economic Surveys}, 7,  145–-197.
    
    \smallskip
    
    \rf Griffiths,W.E. and P.A.A. Beesley (1984), The small-sample properties of some preliminary test estimators in linear model with autocorrelated errors. \textit{Journal of Econometrics}, 25, 49--61.
    
    \smallskip
    
    \rf Hammersley, J. M. and Handscomb, D. C. (1965), Monte carlo methods. Methuen, London.

    \smallskip
    
    \rf Hildreth, C. and J. Lu (1960), Demand relations with autocorrelated disturbances. Technical Bulletin No. 276, Michigan State University.

    \smallskip
    
    \rf Imhof, J.P. (1961), Computing the distribution of quadratic forms in normal variables. \textit{Biometrika}, 48, 419--426.

    \smallskip
    
    \rf Kabaila, P. (2009), The coverage properties of confidence regions after model
    selection. \textit{International Statistical Review}, 77, 405--414.
    
    \smallskip
    
    \rf Kabaila, P., R. Mainzer and D. Farchione (2015), The impact of a Hausman pretest, applied to panel data, on the coverage probability of confidence intervals. \textit{Economics Letters}, 131, 12--15.
    
    \smallskip
    
    \rf Kabaila, P., R. Mainzer and D. Farchione (2017), Conditional assessment of the impact of a Hausman pretest on confidence intervals. \textit{Statistica Neerlandica}, 71, 240--262.
    
    \smallskip
    
    \rf Karmel, P.H., and M. Polasek (1977), Applied statistics for economists, 4th edition. Pitman Australia, Carlton, Victoria.
    
    \smallskip
    
    \rf Katz, D.A. (1982), Econometric theory and applications. Prentice-Hall, Englewood Cliffs NJ.
    
    \smallskip
    
    \rf Kennedy, P. (2008), A guide to econometrics, 6th edition. Blackwell, Malden MA.

    \smallskip
    
    \rf Pokorny, M. (1987), An introduction to econometrics. Blackwell, London.

    \smallskip
    
    \rf Salvatore, D. and D. Reagle (2002), Theory and Problems of Statistics and Econometrics. McGraw-Hill, New York.

    \smallskip
    
    \rf Studenmund, H.A. (1992), Using econometrics, a practical guide, 2nd edition. Harper Collins, New York. 
    
    \rf Studenmund, H.A. (2006), Using econometrics, a practical guide, 5th edition. Pearson, New York.

    \smallskip
    
    \rf Verbeek, M. (2004), A guide to modern econometrics, 2nd edition. John Wiley, London.
    
    \smallskip
    
    \rf Wooldridge, J.M. (2016), Introductory econometrics: a modern approach,
    6th edition. Cengage Learning, Boston, MA.

\end{document}